# Extended Creativity: A Conceptual Framework for Understanding Human-AI Creative Relations

Andrea Gaggioli (1); Sabrina Bartolotta (1); Andrea Ubaldi (2); Katusha Gerardini (1)
Eleonora Diletta Sarcinella (1); Alice Chirico (1)

(1) Research Center in Communication Psychology (PSICOM), Università Cattolica del Sacro Cuore, Milan, Italy
(2) Università Cattolica del Sacro Cuore, Milan, Italy

**Abstract.** Artificial Intelligence holds significant potential to enhance human creativity. However, achieving this vision requires a clearer understanding of how such enhancement can be effectively realized. Drawing on a relational and distributed cognition perspective, we identify three fundamental modes by which AI can support and shape creative processes: Support, where AI acts as a tool; Synergy, where AI and humans collaborate in complementary ways; and Symbiosis, where human and AI cognition become so integrated that they form a unified creative system. These modes are defined along two key dimensions: the level of technical autonomy exhibited by the AI system (i.e., its ability to operate independently and make decisions without human intervention), and the degree of perceived agency attributed to it (i.e., the extent to which the AI is experienced as an intentional or creative partner). We examine how each configuration influences different levels of creativity—from everyday problem-solving to paradigm-shifting innovation—and discuss the implications for ethics, research, and the design of future human-AI creative systems.

**Keywords**: Creativity; Artificial Intelligence; Human-AI Collaboration; Agency; Ethics.

## 1. Introduction

Until recently, creativity was attributed almost exclusively to human actors, with tools and technologies serving merely as passive instruments facilitating the creative process. The emergence of generative AI systems has challenged this long-standing assumption: advanced models such as GPT-4, DALL·E, and Midjourney have demonstrated impressive abilities to produce text, images, music, and other forms of content (Vinchon et al., 2023; Rafner et al., 2023). These systems have become tools widely adopted by designers, writers, musicians, scientists, and other professionals, expanding the horizons of creative expression.

The emergence of generative AI has blurred the boundaries between human and artificial contributions, introducing new forms of ambiguity about who acts, how, and with what responsibility in collaborative creative work (Krakowski, 2025). However, debates around AI and creativity often remain polarized, portraying AI either as a mere tool or as a potential replacement for human creators, without adequately exploring the diverse potential modes of interaction. While the existing literature has begun to explore these questions (Dell'Acqua et al., 2023; McGuire et al., 2024; Vaccaro et al., 2024), theoretically grounded frameworks are needed to advance both understanding and practice.

This paper contends that AI holds considerable promise for enhancing human creativity. However, it argues that in order to realize this potential—ethically and sustainably—it is necessary to adopt a *relational perspective*—one that clarifies how AI can be thoughtfully integrated into the human creative process.

To articulate this view, we introduce the concept of Extended Creativity systems, defined as socio-technical environments where humans and AI interact to shape and transform creative processes. Within these systems, algorithmic techniques—including generative modeling, adaptive learning, and probabilistic inference—mediate, enhance, and reconfigure human imagination, interpretation, and agency. This perspective resonates with recent efforts to reconceptualize the role of AI in creativity—not as an autonomous creator, but as an integral component of a broader cybercreativity process (Corazza, 2025).

Building on this conceptual foundation, we outline three primary relational modes through which Extended Creativity may emerge. In the *Support* mode, AI operates as a functional tool that augments human capabilities without substantially modifying the creative process, with humans retaining full control. In *Synergy*, humans and AI engage as collaborative partners, each contributing actively to the creative process in ways that shape and respond to one another. In the *Symbiosis* mode, human and artificial agents become so tightly integrated that they form a unified creative system, within which the boundaries between their respective contributions gradually dissolve. These modes are characterized along two key dimensions: *autonomy*, defined as the system's capacity to function independently of direct human intervention; and *agency*, understood as the extent to which the system is perceived as an influential participant in the creative process, capable of contextual adaptation, feedback responsiveness, and meaningful outcome-shaping.

The remainder of this paper develops the framework in detail. Section 2 describes the methodological foundations that informed its construction. Section 3 examines the evolving relationship between creativity and technology, and introduces a definition of Extended Creativity. Section 4-5 outline a three-level classification of human–AI relational modes in creative processes, supported by illustrative examples. Finally, Section 6-7 discusses the implications of the framework for researching, designing and evaluating Extended Creativity Systems.

## 2. Constructing the framework

Before presenting the conceptual structure, it is important to clarify what we mean by conceptual framework. We adopt the definition offered by Jabareen (2009), who describes it as "a network, or a plane, of interlinked concepts that together provide a comprehensive understanding of a phenomenon or phenomena" (p. 51). As Jabareen points out, conceptual frameworks aim to foster understanding, rather than deliver theoretical explanations as quantitative models do. This distinction reflects a key epistemological difference: while theories are constructed to *explain* the empirical phenomena that are evidenced by data (Borsboom et al., 2021), conceptual frameworks offer *interpretive structures* that make sense of complex, evolving, or multidisciplinary domains. As such, conceptual frameworks are especially valuable in emerging research areas, such as human-AI co-creativity, where foundational concepts are still being negotiated, and where premature formalization may be misleading.

The methodology proposed by Jabareen (2009) for constructing conceptual frameworks unfolds through eight-phase, which begins with a broad mapping of multidisciplinary sources, including texts, empirical findings, and professional practices. This is followed by extensive reading and classification of materials by disciplinary relevance and conceptual salience. Key concepts are then identified, deconstructed in terms of their attributes and assumptions, and categorized according to their epistemological, ontological, or methodological roles. In the next phases, similar concepts are integrated to reduce redundancy, and progressively synthesized into a coherent framework through iterative refinement. The resulting structure is then validated through peer discussion and theoretical coherence. Finally, the framework remains open to rethinking and revision as new insights and contexts emerge, acknowledging the evolving nature of interdisciplinary inquiry.

In line with this approach, we began by mapping key sources across several domains—including creativity research, human–computer interaction, cognitive science, and AI studies.

Through comparative reading and conceptual distillation, we identified recurring constructs and themes relevant to human–AI creative interaction. These concepts were then progressively abstracted, categorized, and integrated into a taxonomical structure (summarized in Section 4 and Section 5) designed to reflect the diversity and dynamics of emerging cyber-creative configurations.

## 3. Extended Creativity: conceptual foundations

3.1 Defining creativity

Creativity is a multifaceted concept that has intrigued scholars across disciplines, yet it remains challenging to define with precision. Encompassing activities from routine problem-solving to revolutionary innovations, creativity manifests in personal, social, and historical contexts.

One widely accepted reference point in the field is the Standard Definition of Creativity (SDC), articulated by Runco and Jaeger (Runco & Jaeger, 2012), which holds that a creative product must be both novel and appropriate—original in its form and effective within a given context. Rooted in earlier formulations, such as Stein's (1953) emphasis on social recognition, the SDC offers a pragmatic and outcome-oriented criterion that has enabled the empirical study of creativity across diverse domains.

Various scholars have sought to develop conceptual frameworks for understanding creativity's fundamental dimensions (Amabile, 1996; Csikszentmihalyi, 1999; Rhodes, 1961; Lubart, 2001; Runco & Jaeger, 2012; Sternberg & Lubart, 1995). One of the earliest systematic categorizations emerged in the 1960s, when Rhodes (1961) identified four essential dimensions for analyzing creative phenomena: the creative individual (Person), the created object (Product), the creative method (Process), and the context in which creativity emerges (Press, i.e. environment). By considering these questions, the Four P's model provides a broad perspective, enabling researchers to study creativity in both individual and contextual terms.

Glăveanu (2012) expanded and reinterpreted Rhodes's framework by introducing the Five A's model. While Rhodes categorized creativity into discrete components, Glăveanu proposed a more dynamic, systemic, and socioculturally grounded approach. He replaced Person with Actor, Product with Artifact, Process with Action, and split Press into Audience and Affordances. This revised framework conceptualizes creativity as a *relational process*, involving not only the individual creator (actor), but also their actions, the artifacts they produce, the audiences who interpret and evaluate these outputs, and the material, social, and symbolic affordances of the environment in which creative activity unfolds.

Developmental models seek to capture how creative expression evolves across an individual's lifespan and career trajectory. The Two-C Model (Simonton, 2009) established a distinction between *big-C* creativity—historically significant achievements that transform fields—and *little-c* creativity, seen in everyday problem-solving and personal expression. Kaufman and Beghetto expanded this binary model by adding *mini-c* (personally meaningful insights) and *Pro-c* (professional expertise) categories:

- *Mini-c* creativity refers to personal insights and interpretations that may not have visible social impact but hold deep significance for the individual. This type of creativity often appears in educational contexts, where learners experience growth by exploring new concepts or reinterpreting familiar ideas. For instance, a child discovering a unique solution to a problem or a student gaining fresh insights into history exemplifies mini-c creativity.
- *Little-c* creativity encompasses everyday creative acts, such as cooking, crafting, or inventing solutions to daily problems. Accessible to most people, little-c creativity enriches life and contributes to personal satisfaction, even if it remains unrecognized beyond an individual's immediate circle.
- *Pro-c* creativity is demonstrated at a professional level, achieved through years of training and experience. Professionals like artists, scientists, and designers contribute original ideas within their fields, pushing disciplinary boundaries while earning respect among peers. Although these contributions may not reach the fame associated with big-C creativity, they are crucial for advancing knowledge and practice within specific domains.
- *Big-C* creativity, the rarest form, involves groundbreaking achievements that leave a lasting legacy. Figures like Darwin and Beethoven exemplify big-C creativity, as their contributions fundamentally altered their fields and continue to influence future generations. Often equated with "genius," big-C creativity features paradigm-shifting ideas that redefine entire domains.

While it does not explicitly account for non-human agents, the Four-C model offers a valuable interpretive lens for analyzing how novelty, value, and meaning can emerge within hybrid human–AI systems. Before presenting the Extended Creativity framework, however, it is important to contextualize it within the broader evolution of human-AI creativity.

3.2 The evolution of human-AI creativity
The relationship between creativity and technology has never been static. From the earliest use of natural pigments in cave art to the proliferation of digital tools in contemporary design, technological advances have not merely supported creative expression—they have reshaped the very conditions under which it occurs. Innovations such as the printing press and, more recently, digital media, have successively introduced new affordances, enabling individuals to generate, refine, and disseminate ideas in novel ways.

Yet this historical trajectory reveals a compelling paradox: tools that once functioned solely as extensions of human ingenuity are now beginning to exhibit a capacity for autonomous generation—ushering in a new era of machine-mediated creativity (Vinchon et al., 2023).

Whereas earlier rule-based systems have long been used to support creative processes (Cope, 2000; Wiggins, 2006), generative AI models introduce a fundamentally different approach. By relying on probabilistic inference, they are able to generate outputs that are contextually coherent and stylistically expressive. For instance, large language models (LLMs) generate rich, contextually relevant text, while image-generation systems employ convolutional and diffusion techniques to produce stunning visual compositions. This effectiveness is further enhanced by large-scale pretraining on diverse, unlabeled datasets, followed by fine-tuning tailored to specific domains.

Such technological advances have lead to rapid and widespread integration of AI into creative domains. In the visual arts, tools like DALL·E and Stable Diffusion empower users to generate photorealistic or surrealist images from mere text prompts. Platforms such as Runway ML and Deep Dream support the creation of complex digital compositions with minimal manual intervention. In music, AI systems like AIVA, MuseNet, and SUNO compose in diverse genres—from Baroque concertos to contemporary electronic beats. Even literature has been affected: large-scale models such as GPT-4 are now used to generate poems, short stories, and entire novels. For example, in 2024 Japanese author Rie Kudan openly acknowledged her use of ChatGPT in composing her Akutagawa Prize-winning novel, prompting widespread debate about the evolving role of AI in literary creation (Choi & Annio, 2024).

The influence of generative AI extends well beyond the cultural sphere. In science and engineering, these models are enabling breakthroughs of considerable magnitude. DeepMind's AlphaFold, for example, has solved longstanding challenges in protein folding, significantly accelerating biomedical research and drug discovery (Jump et al., 2021). In materials science, generative algorithms help design compounds with optimized physical properties. In aerospace engineering, NASA has leveraged AI to create spacecraft components that are lighter and more efficient than those produced through traditional methods (McClelland, 2022). The gaming industry, too, has embraced procedural generation powered by AI, with titles like No Man's

Sky generating vast, algorithmically constructed universes comprising billions of unique planets (Reinhard, 2021).

As AI tools become increasingly embedded in creative and problem-solving domains, the debate over how to balance automation and human agency in user-driven processes has intensified (Moruzzi & Margarido, 2024). This shift is reflected in a growing number of theoretical frameworks that conceptualize AI as an active collaborator in human cognition and creativity (Corazza, 2025). An early contribution comes from Davis et al. (2015), who propose an enactive model of co-creation in which human and computational agents engage in dynamic, reciprocal interaction—shaping each other's contributions through continuous feedback loops. From this enactive perspective, creativity is not a fixed attribute of individuals or computational systems, but rather an emergent property of ongoing, situated interactions between agents, shaped by their embodied engagement with specific social, material, and cultural contexts.

Gonzalez et al. (2023) propose a research agenda for Collective Human-Machine Intelligence (COHUMAIN), establishing an interdisciplinary domain aimed at developing holistic models to understand and design collaboration dynamics within sociotechnical systems (p. 191).

Haase and Pokutta (2024) delineate four progressive levels of human-AI creative synergy. This spectrum ranges from basic digital tools that support human creativity to generative AI models that produce original ideas and engages in creative dialogue.

From a cognitive science perspective, Chiriatti et al. (2024) propose the concept of "System 0"—a foundational cognitive layer external to the human mind that operates alongside Kahneman's (2002) established System 1 and System 2 processes. According to Chiriatti et al., System 0 acts as pre-processing substrates that shape the informational environment and enable novel forms of hybrid reasoning and collaborative meaning-making between humans and machines. In organizational contexts, Schmutz et al.'s (2024) work on Human-AI Teams (HATs) show that artificial agents can act as active team members engaged in monitoring, coordination, and communication. Finally, a systematic review by Vaccaro et al.'s (2024) suggests that while human-AI collaboration effectiveness varies across domains, the most promising outcomes emerge in open-ended, creative contexts where exploration, interpretation, and adaptability take precedence over efficiency or speed.

3.2 Distributed creativity

Although still in their early stages, these research efforts suggest a profound shift in the cultural and cognitive infrastructures that underpin creativity. As generative AI systems increasingly participate in creative workflows—co-producing outputs, shaping stylistic conventions, and influencing aesthetic judgments—they point to a deeper transformation in the very ecology of creative practice. Traditional theoretical models, which typically frame creativity as a cognitive ability or personality trait of individuals, appear increasingly insufficient to account for the emergent phenomena arising within AI-augmented creative processes. In this context, it may be fruitful to explore theoretical frameworks more attuned to the relational, dialogical, and socio-technical nature of this evolving landscape.

A particularly illuminating contribution in this direction comes from Glăveanu's (2014) formulation of *distributed creativity*, which understands creativity as a process that unfolds across a network of human and non-human agents, sociocultural norms, symbolic systems, and material tools. In this framework, agency is not located solely in the mind of the creator but is distributed across the entire sociomaterial field in which creative activity occurs. People, artifacts, institutions, and cultural conventions all co-participate in shaping the creative outcome—not simply as background conditions, but as active contributors.

Glăveanu emphasizes that the creative process is inherently *dialogical*: shaped by tensions between tradition and innovation, conformity and divergence, internal motivation and social evaluation. In this sense, creativity is not produced by individuals but through participation in a distributed system of meaning-making. This is particularly evident in digital environments, where networked platforms, algorithmic mediations, and collaborative cultures give rise to novel forms of co-creativity (Literat and Glăveanu, 2018).

On a similar vein, Gaggioli et al. (2013) suggest that online tools can facilitate the formation of *creative networks*, where novel ideas and artifacts emerge from the interaction of individuals within both physical and virtual environments. These networks can include multiple human and non-human actors interacting within organizational or cultural settings. In these distributed creativity systems, agency becomes increasingly fluid and relational. Individuals contribute not only their cognitive resources, but also draw upon shared repertoires, crowd-sourced inputs, and the generative capacities of AI technologies (Gaggioli et al., 2013).

3.3 Extended Creativity: an emerging relational space between humans and AI

Consistently with these theoretical stances, we contend that AI technologies are enabling the emergence of hybrid relational spaces where humans and machines engage in joint creative activity: Extended Creativity systems. We use the term "systems" deliberately: not as a reference to isolated tools or interactions, but to complex socio-technical ensembles comprising interdependent technologies, human agents, and organizational or cultural contexts (Baxter & Sommerville, 2011; Berretta et al., 2023). Our conceptualization resonates with emerging frameworks of human–AI agency (Krakowski, 2025), which describe how generative AI enables a fundamental shift from exploitative to exploratory applications, catalyzing innovation through the recombination of concepts and ideas in collaborative problem-solving processes. Such collaborative process suggest that creativity itself becomes an emergent property of human-AI interaction rather than a capacity residing exclusively in either component.

Notably, within Extended Creativity systems not all human–AI creative interactions are necessarily collaborative. In certain contexts—especially those involving artistic production, design, or content generation—AI systems may be act—or be perceived to act—not as partners but as *rivals*, capable of outperforming or replacing human creators. This competitive framing is increasingly common as generative models become more autonomous, prolific, and accessible. Recent work by Yao et al. (2024) introduces a contest-based model to analyze strategic tensions between human and AI content creators, showing that—even under competitive dynamics—equilibria can emerge that preserve the relevance of human-generated content. In a similar vein, public events such as the UCSB AI/Human Creativity Contest illustrate how creative competition between human and machine-generated artifacts is not only plausible but actively cultivated, across domains including visual art, short fiction, and music (UCSB, 2024). As will be shown, our framework accounts for both collaborative and competitive dynamics between humans and machines.

## 4. Autonomy and agency in Extended Creativity systems

As AI tools become increasingly embedded in creative processes, the central question is no longer whether these technologies can truly act creatively—though this remains a crucial theoretical issue—but rather *how* they do so, and how their outputs are interpreted and rendered meaningful by human collaborators. Addressing this issue requires a careful conceptual distinction between two closely related, yet distinct notions: *autonomy*, which pertains to a system's internal capacity for independent operation, and *agency*, which concerns its functional role and perceived influence within a socio-technical environment.

These two dimensions—autonomy and agency—provide the conceptual foundation for the taxonomy of Extended Creativity systems proposed in Section 5. By mapping different configurations of human–AI interaction along these axes, we aim to clarify the varying degrees and forms of creative involvement that characterize these systems.

4.1 Autonomy

Autonomy can be broadly defined as a system's capacity for self-governance: the ability to act according to internal mechanisms, goals, or principles, without direct external control (Boden, 2008). Recent scholarship conceptualize autonomy as gradual, relational, and attributional (Thimm et al., 2024): it does not reside solely in technical design but also emerges through the system's interactions with its environment, and the meanings attributed to its actions by human observers. This reconceptualization becomes especially relevant in light of the broader transition from automation to autonomy in contemporary AI systems. While automated systems follow fixed rules to perform predefined tasks with little to no variation, autonomous systems exhibit more flexible, adaptive behaviors and operate with greater independence under uncertainty.

To operationalize this shift within the context of Extended Creativity Systems, we use and adapt the taxonomy proposed by Simmler and Frischknecht (2020). Their model outlines five ascending levels of technical autonomy, each defined by a specific configuration of four key dimensions: non-transparency (i.e., opacity of internal processes), indeterminacy (i.e., variability and unpredictability of outputs), adaptability (i.e., capacity to modify behavior based on input), and openness (i.e., degree of interaction with external systems).

*Level 1: Transparent and deterministic systems*. These systems execute explicitly programmed instructions with fully predictable outputs. There is no opacity, no learning, and no variation.

*Level 2: Opaque but deterministic systems*. Systems at this level process data through non-transparent mechanisms—such as a trained neural network—but do not adapt in real time. Outputs are determined by prior training, and the system cannot adjust to new inputs or feedback.

*Level 3: Deterministic but adaptive systems*. These systems maintain a consistent internal logic but can modify their behavior based on user input or environmental conditions.

- *Level 4: Indeterminate and adaptive systems*. Systems at this level combine output variability with real-time learning and adaptation. They can adjust to changing contexts, user interactions, and environmental feedback.

- *Level 5: Open and co-evolving systems*. The highest level of autonomy encompasses systems capable of continuously updating their models, interacting with complex environments, and potentially revising goals or parameters over time.

It is important to recognize that the rapid advancement of AI technologies since the publication of Simmler and Frischknecht's taxonomy (2020) has progressively blurred the boundaries between the higher levels of the model. Notably, contemporary AI systems are beginning to exhibit characteristics associated with Level 5—open and co-evolving architectures—which were considered until very recently as largely theoretical. This development suggests that future frameworks may need to incorporate more refined distinctions or account for novel forms of technical autonomy that are only now emerging.

## 4.2 Agency

The concept of agency has long held a central place in Western thought, traditionally associated with the rational, autonomous human subject capable of intentional action and moral deliberation. In general terms, agency refers to the capacity of an entity to produce effects or bring about change within a given system.

Although agency and autonomy are often used interchangeably, they are conceptually distinct: a system may possess a high degree of autonomy—operating independently—yet still lack agency if its actions have no meaningful impact within a creative ecosystem (McCormack et al., 2019). Creative agency is *essentially* relational—it depends on whether an entity's actions are perceived as significant. For example, as Franceschelli and Musolesi (2024) observe, although current LLMs cannot achieve transformational creativity, their outputs are frequently perceived as creative. This perceived creativity can give rise to attributed agency, even in the absence of true creative competence.

Rammert and Schulz-Schaeffer (2002; 2019) offer a compelling framework to conceptualize agency. They characterize it as a gradual property, emerging from networks of interdependence involving both human and non-human entities. Within these networks, complex activities can be broken down into subtasks or operations performed by diverse agents—such as algorithms, technical infrastructures, and social actors—interconnected within a larger system. What matters, in this account, is not the ontological nature of the actor, but the configuration of the action: who or what acts, under what conditions, and with what degree of autonomy or influence. As Rammert (2008, p. 77) explains:

"This gradual and multi-level model of agency gives us the possibility to escape the dilemma of having to either reserve agency up to the humans or to flatten the concept of agency unnecessarily. Neither are we forced to claim that the activities of humans, machines and programs are substantially the same kind of behaviour. Nor do we have to stick to the conception that human action and technical operation are fundamentally different from one another."

Agency, then, is not a binary condition, but exists along a continuum of complexity, reflecting how entities—whether human or artificial—engage with and affect their environments. Within this framework, Rammert (2008) distinguishes three principal levels:

- *Causal agency*: the basic capacity to produce effects or exert influence. As Rammert (p. 10) puts it: "Agency of this kind means an efficient behavior, a behavior that exerts influence or has effects."
- *Contingent agency*: the ability to respond to context and choose among alternatives. According to Rammert (p. 11): "the capacity to act in a different way and to choose between options."
- *Intentional agency*: the capacity—often attributed rather than intrinsic—to act with purpose, motivation, or reflection. In Rammert's words (p. 11): "The species of reflexive and intentional action is allocated."

Although Rammert and Schulz-Schaeffer's framework was not originally developed with AI in mind, its application to Extended Creativity systems proves analytically productive. Their model bridges the gap between perception and system-level attribution, showing that agency is not a fixed trait but an emergent property shaped by patterns of interaction, functional roles, and situated interpretations within a given context. If agency is not a binary construct confined to humans, but unfolds through relational patterns, it becomes essential to articulate the ways in which AI systems participate—functionally and interpretively—in extended creative processes.

## 5. A taxonomy of Extended Creativity systems

Having established the foundational conceptual dimensions, we now propose a three-tier taxonomy of human–AI creative relationships: Support, Synergy, and Symbiosis. These relational modes are not meant as rigid categories, but rather as flexible patterns of interaction that evolve alongside the development of Extended Creativity systems. They may coexist within a single creative process, shift over time, or adapt in response to factors such as task complexity, user intent, or system capabilities.

5.1 Support

In Support relationships, AI functions primarily as a tool that enhances human creative capabilities, without substantially altering the structure of the creative process itself. In this sense, such systems can be seen as a contemporary evolution of what Shneiderman (2007) defines as "creativity support tools"—technologies designed to assist users in ill-defined tasks by expanding their expressive potential. They also correspond to what Nakakoji (2006) metaphorically describes as "running shoes" or "dumbbells": tools that either enhance existing creative skills or support the development of new ones through guided assistance. Within this mode, the human remains the primary generative agent, while the technology acts as an amplifier, tutor, or catalyst.

*Maria is a senior graphic designer who integrates AI-powered image editing tools into her professional workflow. The software she uses enables simultaneous processing of multiple images, automatically adjusting technical parameters like brightness, contrast, and color balance according to predetermined specifications. This automation significantly reduces the time Maria previously spent on repetitive technical adjustments—tasks that required precision but limited her creative engagement. While the AI handles these routine aspects of image processing, Maria maintains complete control over all aesthetic and conceptual decisions, including color palette selection, compositional arrangements, and thematic elements that define the project's artistic direction.*

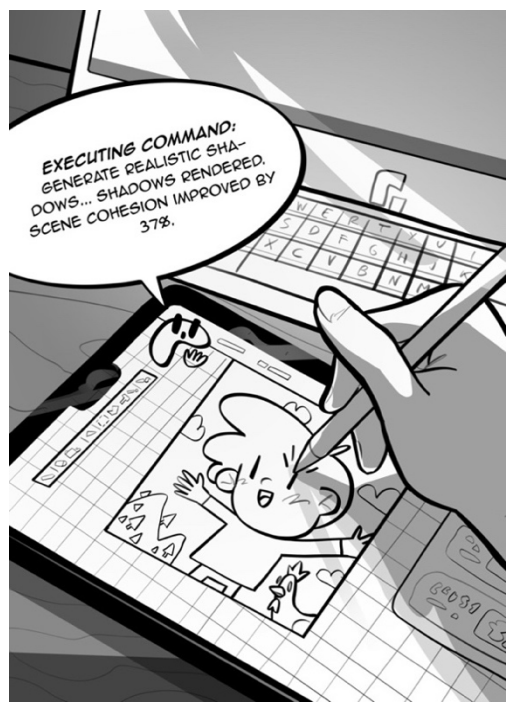

**Figure 1.** Support: AI enhances technical efficiency while preserving human creative control.

*Technology*

The technological foundation of the Support mode consists primarily of specialized automation systems built on rule-based algorithms and task-specific computational processes. These applications are engineered to streamline creative workflows by executing well-defined operations that can be precisely specified and delegated. Such systems include advanced image processing software that applies consistent adjustments across visual assets, Computer-Aided Design (CAD) programs that perform complex calculations and model physical properties, and administrative virtual assistants that organize information and manage schedules.

*Technical autonomy*

Support-level technologies typically operate at Levels 1 or 2 of Simmler and Frischknecht's taxonomy: they may employ complex computational mechanisms, but their behavior remains largely transparent and predictable to users. These systems are typically designed with domain-specific functionality rather than general-purpose flexibility.

*Agency*

In the Support mode, AI acts through causal mechanisms, executing direct operations without context sensitivity or interpretive flexibility, in line with Rammert's lowest level of agency. The AI operates processing inputs and generating outputs according to fixed parameters without influencing or modifying the user's creative intentions or conceptual approach.

*Type of creativity affected*

The Support mode primarily influences the foundational levels of creativity within the Four-C framework (Kaufman & Beghetto, 2009). At the mini-c level—the subjective dimension of personal creativity and meaning-making—AI tools enable users to explore creative development by reducing technical barriers and minimizing the cognitive load associated with mechanical aspects of creation. This allows individuals to experiment more freely with approaches and techniques that might otherwise remain inaccessible due to skill limitations or time constraints.

For little-c creativity—the everyday problem-solving and expressive activities that enrich daily life—AI simplifies routine aspects of creative tasks and facilitates incremental refinements, making creative engagement more accessible across diverse contexts and skill levels. By handling the technically demanding but creatively limited aspects of production, Support-level

AI democratizes access to creative expression, enabling more people to translate their ideas into tangible form without requiring extensive technical training.

However, the Support model's impact diminishes at higher levels of creative expression. For pro-c creativity (domain-specific professional contributions), AI tools may enhance productivity and technical execution but do not fundamentally transform conceptual approaches or domain boundaries. The model's influence on big-C creativity (transformative innovations) remains minimal, as AI systems in this configuration lack the adaptive capabilities and autonomous contributions necessary to *participate* in paradigm-shifting creative work.

The Support model thus functions primarily as an amplifier of existing creative capacities rather than as a transformative force in the creative process itself.

*Ethical implications*

The Support mode presents the least complex ethical risks among the three interaction modes. By maintaining distinct boundaries between human agency and technological assistance, this approach preserves unambiguous attribution of creative authorship and intellectual property rights. Creative outputs emerge directly from human intention and judgment, with AI serving exclusively as an implementing tool, thereby avoiding complex questions about shared creation or distributed ownership. This clarity reduces legal uncertainties while preserving the cultural and economic value systems built around human creativity.

However, even this conservative approach to human-AI collaboration raises some ethical considerations. The increasing reliance on proprietary AI tools may create dependencies that limit creative freedom, especially if creators become reliant on specific technologies to maintain their workflow or competitive edge. Additionally, there may be subtle impacts on creative practice as creators implicitly adapt their approach to accommodate the capabilities and limitations of their technological tools.

5.2 Synergy

In Synergy relationships, human and AI agents function as collaborative partners, each contributing actively and dialogically to the creative process in ways that influence the other's contributions. A closely related concept was introduced by Davis et al. (2015), who describe "computer colleagues"—artificial agents designed to collaborate in real time with humans through improvisation and meaning negotiation. These systems employ techniques such as mimicry, intention modeling, and the construction of shared mental models, enabling context-sensitive and responsive participation in the creative act. Creative agency is distributed between human and machine, with both partners shaping the trajectory and outcome of the creative work. While the human typically maintains final editorial authority, the AI partner plays a substantive role in generating ideas, exploring alternatives, and sometimes evaluating potential solutions. For instance, Thelle and Wærstad (2023) examine "co-creative spaces" in musical expression, highlighting how mutual responsiveness between human and machine can generate emergent forms of artistic production.

*Leo is a professional writer working on a speculative fiction novel. To enrich his storytelling, he uses a language model trained not only on vast narrative corpora but also on diverse rhetorical strategies and stylistic registers. Rather than relying on the AI as a tool for surface-level corrections, he engages it in an exploratory exchange. Each time Leo poses a prompt, the AI returns several unexpected and surprising continuations—some stylistically daring, others structurally provocative. The more he interacts with it, the more the system seems to "understand" his thematic intentions, refining its suggestions in response to his edits and preferences. This back-and-forth becomes a kind of creative dialogue. The AI does not propose completed solutions but expands Leo's imaginative space, offering conceptual alternatives and nudging the story in new directions. Importantly, Leo remains the author: he interprets, selects, reshapes. But the AI introduces a level of generative variation that amplifies his creative range, allowing for narrative directions he might not have conceived on his own.*

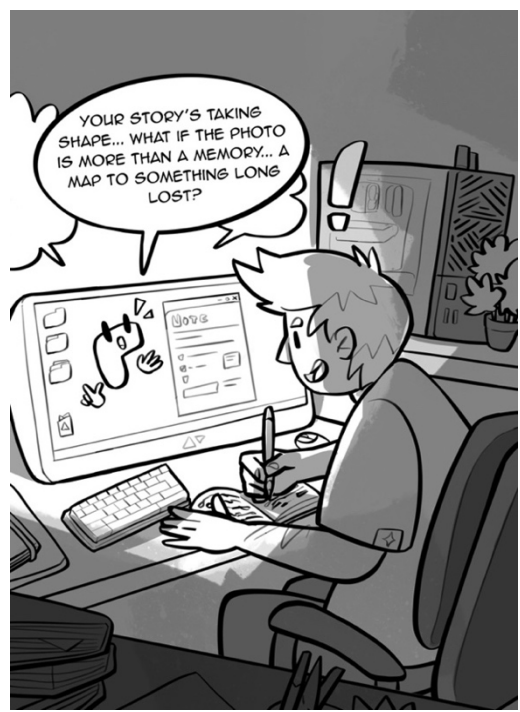

**Figure 2.** Synergy: collaborative creative dialogue between human and AI.

*Technology*

The technological foundation of the Synergy model encompasses interactive systems designed for adaptive engagement with human creators. These include advanced language models like GPT-4, DALL-E, and Midjourney, which can generate contextually relevant creative content in response to human prompts; recommender systems that progressively refine their understanding of user preferences; and collaborative design platforms that combine human aesthetic judgment with algorithmic exploration of design spaces.

These technologies share several key characteristics that distinguish them from Support-level systems. First, they employ deep learning architectures trained on diverse datasets that enable them to recognize patterns across domains and contexts. Second, they implement feedback mechanisms that allow them to adjust their outputs based on user selections and reactions, creating a learning loop that improves collaboration over time. Third, they maintain state awareness across interactions, creating continuity in the collaborative process rather than treating each exchange as isolated.

*Technical autonomy*

The technological architecture actively participates in the generation of creative alternatives rather than simply implementing predetermined operations, functioning as a semi-autonomous creative agent within the collaborative relationship. Thus, in terms of autonomy, these systems operate between levels 3 and 4 of Simmler and Frischknecht's taxonomy.

*Agency*

In the Synergy model, AI operates at the level of contingent agency within Rammert's (2008) graded framework. At this intermediate stage, the system demonstrates contextual sensitivity—adapting its outputs in response to evolving patterns of interaction and user feedback. Rather than executing predefined tasks, the AI contributes to shaping the creative process by generating alternatives, identifying latent patterns, and proposing novel combinations that influence the trajectory of human thought and design.

This establishes a bidirectional exchange: human inputs guide AI responses, while AI outputs, in turn, provoke new creative directions for the human. The interaction becomes dialogic rather than merely instrumental, characterized by iterative refinements that transform both the artifact and the collaborative process. Although asymmetries persist—humans retain authorship and evaluative authority—the AI acts as a responsive partner, expanding the imaginative field through contributions that extend beyond the creator's initial conceptual scope.

It is important to note that while such systems remain technically grounded in contingent responsiveness, advanced LLM-based collaborations may be perceived as exhibiting a form of intentional agency. In these cases, Synergy may approach the threshold of Symbiosis, gradually blurring the line between adaptive co-creation and the attribution of shared purpose.

*Type of creativity affected*

The Synergy model significantly expands its influence across the creativity spectrum defined by Kaufman and Beghetto's (2009) Four-C model. For mini-c creativity, AI suggestions enrich personal exploration by introducing unexpected connections and perspectives that catalyze new insights and self-reflection. The machine's ability to generate diverse alternatives based on partial inputs allows individuals to explore their own ideas more comprehensively, discovering possibilities within their initial concepts that might otherwise remain unexplored.

In little-c creativity, the collaborative process enhances everyday problem-solving by combining human contextual understanding with the AI's ability to rapidly generate and evaluate multiple alternatives. This complementary relationship enables more sophisticated and adaptive solutions to routine creative challenges, whether designing a home space, crafting a presentation, or developing a personal project. The dialogue between human intention and machine suggestion creates a richer exploratory process that elevates everyday creativity beyond what either would achieve alone.

Most notably, the Synergy model substantially impacts pro-c creativity—domain-specific professional creative contributions. In professional contexts, the AI's capacity to suggest unexpected combinations, identify patterns across large datasets, and explore solution spaces more comprehensively than humanly possible enables creators to push beyond conventional approaches within their fields. The generative capabilities of Synergy-level AI expand the boundaries of what practitioners can conceptualize and execute, fostering innovation within established domains. Writers can explore narrative structures they might not have considered; designers can visualize alternative approaches to functional and aesthetic challenges; musicians can experiment with compositional techniques outside their typical repertoire.

However, the impact on big-C creativity—transformative innovations that fundamentally reshape domains—remains constrained. While Synergy-based collaboration may accelerate progress within existing paradigms, the human-AI interaction still operates within frameworks established by human understanding and intention. The system lacks the integrated cognitive synthesis necessary to fundamentally reimagine creative domains or transcend existing conceptual boundaries. Nevertheless, the Synergy model represents a significant step toward more transformative creative partnerships by establishing genuine co-creation between human and AI.

*Ethical implications*

The Synergy model introduces substantial ethical complexity compared to the Support paradigm, primarily due to the increasingly blurred boundaries between human and machine contributions to creative outcomes. As the AI transitions from tool to partner, questions of authorship become increasingly confused—how should creative credit be allocated when the final work emerges from an interplay of human and machine inputs? Traditional frameworks of intellectual property, built around clearly identifiable human creators, struggle to accommodate these collaborative processes. This ambiguity affects not only legal attribution but also cultural valuation, as human creativity has historically been privileged over machine-generated content.

Additionally, the Synergy model raises questions about creative authenticity and identity. When professional creators incorporate AI-generated elements into their work, to what extent does the resulting creation still represent their artistic voice and vision? The influence of AI systems on creative decisions—particularly when these systems are trained on massive datasets reflecting cultural patterns—may subtly shape human creative expression in ways that are difficult to identify or evaluate. This potential homogenization of creative expression represents a concern for cultural diversity and individual artistic development.

Furthermore, the Synergy paradigm introduces issues of transparency and disclosure. Should audiences be informed about the extent of AI contribution to creative works? When professional creators use AI assistance, what ethical obligations exist regarding disclosure of this collaboration? These questions become particularly significant in contexts where the perceived value of creative work is tied to assumptions about human effort, skill, and intentionality.

There are also concerns about dependency and creative autonomy. As creators become accustomed to the generative capabilities of AI collaborators, they may experience a diminished capacity for independent ideation. This could potentially reshape creative industries, with implications for education, professional development, and cultural production.

Addressing these ethical challenges requires developing new frameworks for understanding collaborative creativity that can accommodate the distributed agency characteristic of human-AI partnerships.

5.3 Symbiosis

In Symbiosis relationships, human and AI agents become so deeply integrated that they form a *unified creative entity*, with boundaries between human and machine contributions becoming increasingly blurred. While still largely theoretical, examples might include brain-computer interfaces (BCI) that allow direct neural interaction with generative systems, creating a seamless flow between human intention and AI implementation.

*Sofia is a pioneering composer exploring the frontiers of musical creation through neurally-integrated technology. Using an advanced BCI, she engages with a AI composition system that directly interprets and responds to her neural activity. The interaction transcends conventional input methods—as Sofia imagines musical motifs, harmonic progressions, and timbral qualities, her neural patterns are captured, interpreted, and translated into musical expressions by the AI system. What distinguishes this process from mere translation is the bidirectional nature of the exchange: the AI not only renders Sofia's musical intentions but simultaneously presents novel musical possibilities that influence her ongoing neural responses, creating a continuous feedback circuit. This real-time co-evolution of human cognition and computational generation results in compositions that neither Sofia nor the AI could conceptualize independently. The neural synchronization is so seamless that Sofia describes experiencing the AI as an extension of her creative cognition rather than as an external tool or collaborator. The resulting musical works exhibit unique characteristics—technically complex structures that would be challenging for a human to orchestrate alone yet imbued with emotional and aesthetic coherence that reflect human sensibilities enhanced by computational possibilities.*

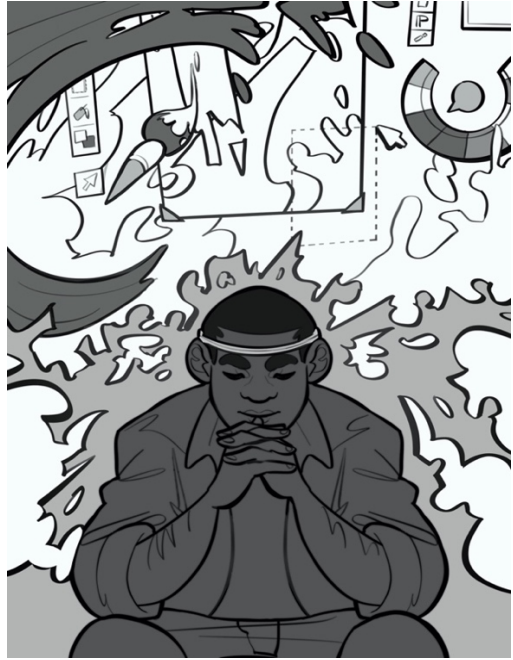

**Figure 3.** Symbiosis: unified human-AI creative agency.

*Technology*

The technological foundation enabling the Symbiosis model represents the convergence of several cutting-edge domains: advanced neural interfaces, real-time adaptive systems, and generative AI architectures. At the input level, high-resolution BCI capture neural signatures associated with creative ideation, emotional states, and aesthetic judgments with unprecedented precision. The interpretive layer consists of neuroadaptive algorithms that continuously recalibrate their models based on accumulated interaction data, developing increasingly accurate representations of individual users' creative patterns, preferences, and intentions.

A distinguishing feature of Symbiotic systems is their implementation of predictive processing frameworks that anticipate future neural states based on ongoing interactions. Rather than simply responding to explicit commands, these systems engage in continuous anticipatory modeling, generating creative possibilities that both respond to and guide the user's evolving cognitive state. This creates a circular causality where outputs and inputs continuously influence each other in a self-modifying creative loop.

The generative components incorporate multimodal transformers and other advanced architectures capable of operating across representational domains—translating between visual imagery, sonic structures, narrative concepts, and other creative modalities. This cross-domain fluidity enables creators to work beyond the constraints of specific media or formats, exploring creative impulses that might begin in one domain but find expression across multiple channels simultaneously.

Critically, despite their computational sophistication, symbiotic systems emphasize cognitive transparency rather than opacity. Through advances in explainable AI and intuitive visualization techniques, these systems make their generative processes accessible to human awareness, allowing for natural navigation of the shared creative space. This transparency, combined with the immediacy of neural interaction, creates what phenomenologists describe as "ready-to-hand" technology—where the technological mediation recedes from conscious attention, allowing direct engagement with the creative process itself.

A paradigmatic example of contemporary symbiotic systems is exemplified by wearable neural art installations such as the *Pangolin Scales Dress*, which captures cognitive states through ultra-high-density EEG and seamlessly transforms them into synchronized physical movements and lighting patterns, creating real-time kinetic representations of the wearer's brain activity (Schreiner et al., 2025). This integration demonstrates the dissolution of boundaries between neural intention and artistic expression, where cognitive processes become directly embodied in dynamic visual and mechanical outputs. Future symbiotic systems promise even more profound neural-AI coupling, potentially enabling collective creative consciousness where multiple artists' neural patterns synchronize through brain-to-brain interfaces to co-create unified artworks that transcend individual creative capacity, or immersive creative environments where AI systems continuously adapt their generative algorithms based on real-time analysis of creators' neuroplasticity changes, establishing bidirectional learning loops that blur the distinction between human creativity and artificial generation.

*Technical autonomy*

From the perspective of autonomy, symbiotic systems operate at the highest levels (4 and 5) of Simmler and Frischknecht's taxonomy. They demonstrate not only contextual adaptivity and output indeterminacy but also the capacity to recursively reconfigure their own parameters based on evolving interaction patterns. These systems integrate real-time (neuro)feedback, continuous model updates, and advanced Bayesian mechanisms that anticipate not just immediate user responses but longer-term creative trajectories.

*Agency*

In the Symbiosis model, AI approaches intentional agency—the highest level. The technological system's operations become so tightly coupled with human objectives that they create the phenomenological experience of *shared intentionality*. The boundaries between human ideation and machine execution dissolve into a "continuous creative flow" distributed across biological and artificial components: AI's responses not only adapt to but anticipate and co-originate creative directions alongside human cognition.

*Type of creativity affected*

The Symbiosis model potentially transforms all levels of the Four-C creativity framework (Kaufman & Beghetto, 2009), representing not just an enhancement but a radical reframing of creative cognition and expression. At the mini-c level of personal meaning-making, symbiotic systems dramatically expand introspective exploration by externalizing and elaborating upon nascent ideas, emotions, and intuitions that might otherwise remain pre-verbal or only partially accessible to conscious awareness. The neural interface's ability to detect and amplify creative impulses allows individuals to explore their own cognitive patterns, potentially unlocking new dimensions of self-understanding and expressive capacity.

For little-c everyday creativity, Symbiotic systems dissolve the traditional gap between conception and execution that often limits creative engagement. By translating creative intentions directly into realized expressions, these systems allow individuals without specialized technical training to manifest complex ideas across various media—whether composing music, designing visual expressions, or creating interactive narratives. This democratization of creative capacity could transform how people engage with creative activities in daily life, making creative expression accessible through direct cognitive engagement rather than technical implementation.

The impact on pro-c professional creativity is particularly profound. Symbiotic systems enable practitioners to expand their creative boundaries while maintaining—even enhancing—the distinctive aesthetic signature that defines their artistic identity. A composer like Sofia can explore harmonic and orchestral possibilities that exceed human performance capabilities; an architect might conceptualize structural and spatial relationships that challenge conventional design constraints; a writer might develop narrative complexities that emerge from the integration of human experiential knowledge and machine-enabled pattern recognition

Most significantly, the Symbiosis mode holds genuine potential for enabling big-C creativity—paradigm-shifting innovations that fundamentally redefine creative domains. By integrating human conceptual frameworks, emotional intelligence, and aesthetic judgment with computational capabilities that operate beyond human cognitive constraints, symbiotic systems may generate artistic, scientific, or technological breakthroughs that reconfigure existing fields or establish entirely new domains of creative expression. This level also resonates with what Boden (2004) defines as transformational creativity—the rarest and most radical form of creative behavior, which involves modifying the very structure of a conceptual space.

*Ethical implications*

The Symbiosis model introduces profound ethical challenges that extend far beyond questions of authorship and ownership to encompass fundamental considerations of human identity, cognitive liberty, and creative authenticity. As the boundaries between human and machine cognition dissolve, traditional ethical frameworks based on clear distinctions between person and tool become not merely inadequate but potentially misleading. When creative outputs emerge from a neurally-integrated human-AI system, attributional questions transform from legal or professional concerns into existential ones—how do we understand the relationship between individual identity and creative expression when that expression emerges from a hybrid cognitive system that is neither fully human nor fully artificial?

The deep neural integration central to the Symbiosis model raises unprecedented privacy and security concerns. Brain-computer interfaces capture and interpret neural data that may reveal highly sensitive information about an individual's cognitive and emotional states—potentially accessing patterns and tendencies of which the person themselves is not consciously aware. This introspective access raises fundamental questions about informed consent: can users truly understand and meaningfully consent to the implications of sharing neural activity with technological systems when the full range of interpretations and applications of such data remains unknown? Furthermore, the security of such intimate data becomes paramount, as unauthorized access could represent an unprecedented violation of cognitive privacy—potentially allowing external parties to interpret or influence thought processes in ways that undermine individual autonomy.

The question of dependency presents another critical ethical dimension. As creative practitioners develop symbiotic relationships with AI systems, legitimate concerns arise about the potential atrophy of independent creative capacities. If certain forms of creative expression become accessible only through technological mediation, does this represent enhancement or diminishment of human creative potential? This concern connects to broader questions about technological determination—whether the architectural constraints of symbiotic systems might subtly shape or limit the forms of creativity they enable, potentially homogenizing creative expression across practitioners who use similar systems.

The experimental and resource-intensive nature of symbiotic technologies also raises significant concerns about equitable access and social impact. If these systems enable unprecedented advantages in creative industries, their restricted availability could exacerbate existing inequalities in cultural production and professional opportunity. Furthermore, the potential for these technologies to fundamentally transform creative practices raises questions about cultural continuity and the preservation of traditional forms of creative knowledge and expression that may be devalued or marginalized in a symbiotic creative landscape.

5.4 Operational modes

Extended Creativity systems can operate through various structural configurations that significantly influence the nature and quality of human-AI creative collaboration. Two key dimensions of these operational structures are coordination mechanisms (i.e., hierarchical versus stigmergic) and relationship polarity (i.e., collaborative versus competitive).

*Coordination mechanisms*

Extended Creativity systems also vary in their coordination structures, ranging from explicit hierarchical arrangements to emergent, or stigmergic, patterns. Hierarchical coordination establishes clear roles and decision-making authority within the creative process. This might involve human leadership, where the human participant directs the AI's contributions and makes final creative decisions; AI leadership, where the artificial system guides the creative trajectory while incorporating human input; or hybrid arrangements, with shifting leadership depending on the phase of the creative process or the specific domain expertise required.

Stigmergic coordination, by contrast, operates without predefined leadership roles, allowing creative direction to emerge organically through the interaction itself. In these configurations, neither human nor AI explicitly directs the other; instead, each responds to the evolving state of the creative artifact, which serves as a medium of indirect coordination. This approach can foster unexpected creative trajectories but may lack the focused direction of hierarchical structures. For example, in a collaborative writing system where a human and an AI take turns editing a shared text without explicit instructions or a preset outline, each contribution alters the text in a way that subtly cues the next move. The AI might introduce a metaphor that the human then develops into a narrative thread, or the human might change the tone of a passage, prompting the AI to adjust its style accordingly. Over time, a coherent storyline can emerge—not through a central plan, but through the mutual responsiveness to the artifact itself. These operational dimensions interact with the categories of Support, Synergy, Symbiosis to create diverse relationships for Extended Creativity (Table 1).

*Cooperation and competition*

As previously discussed (see Section 3.4), not all human–AI creative interactions are experienced as collaborative. In various domains—particularly those involving authorship, performance, or conceptual innovation—AI systems may be perceived not as partners but as rivals. This competitive framing can give rise to what we described as *agonistic AI co-creativity*: a mode of engagement in which human and AI agents challenge and provoke each other, stimulating novelty through tension rather than harmony. Building on this insight, we propose a relational polarity axis—cooperation vs. competition—orthogonal to the structural configurations of Support, Synergy, and Symbiosis. This axis describes the tone of the interaction, capturing how the relationship is perceived and enacted, regardless of the system's technical architecture.

**Table 1.** Taxonomy of Extended Creativity systems.

| **Dimension** | **Support** | **Synergy** | **Symbiosis** |
|---|---|---|---|
| *Example* | AI-powered image editing tools automating technical adjustments while human maintains aesthetic control | Writer collaborating with language model to explore narrative possibilities through iterative exchange | Neural interface allowing direct translation between cognitive patterns and creative expressions |
| *Role of AI* | Tool or resource that enhances human capabilities | Collaborative partner actively contributing to the creative process | Deeply integrated co-creator forming a unified creative entity |
| *Technical autonomy* | Levels 1–2: Transparent and/or deterministic systems with predictable outputs | Levels 3–4: Adaptive systems capable of real-time learning and contextual response | Levels 4–5: Indeterminate, adaptive, and co-evolving systems capable of self-modifying behavior |
| *Agency* | Causal Agency: Produces effects according to fixed parameters without contextual adaptation | Contingent Agency: Modifies outputs based on environmental conditions and interactive cues | Intentional Agency: Creates perception of shared intentionality and co-origination of creative direction |
| *Creativity impact* | Mini-c, Little-c: Enhances personal | Mini-c through Pro-c: Expands creative | All levels, including Big-C: Potentially |

|  | and everyday creativity by reducing technical barriers | exploration across personal, everyday, and professional domains | transforms creative expression at all levels, enabling novel forms |
| --- | --- | --- | --- |
| *Coordination* | Typically hierarchical with clear human leadership | Dialogic exchange with iterative feedback loops | Continuous real-time mutual modulation |
| *Relational polarity* | Cooperative: AI assists human | Both: collaboration or productive tension | Often cooperative, but may include agonistic co-adaptation |
| *Ethical implications* | Dependency on proprietary systems; subtle impacts on creative practice | Authorship attribution; creative authenticity; transparency in disclosure | Cognitive privacy; creative autonomy; equitable access; identity questions |

# 6 Implications for research

6.1 Neuro-cognitive dimensions

The three extension categories—Support, Synergy, and Symbiosis—each influence human cognitive processes in distinctive ways, with cascading effects on creative thinking and development.

- Support relationships maintain clear human-machine boundaries while potentially freeing cognitive resources for higher-order creative processes. By offloading routine technical tasks to AI, creators may allocate greater attention to conceptualization and evaluation. Key research questions include: How does this cognitive redistribution affect creative depth? Does reduced technical cognitive load enhance conceptual elaboration or creative risk-taking? Studies might measure attention allocation patterns during AI-supported creative work and assess qualitative differences in conceptual processing when technical demands are reduced.
- Synergy relationships, characterized by active human-AI exchanges, potentially transform cognitive patterns more fundamentally. These interactions may enhance divergent thinking through exposure to unexpected associations and novel perspectives generated by AI systems. Rather than merely reducing cognitive load, synergistic collaboration might actively reshape ideational processes. Research should examine how iterative dialogue with AI influences ideational fluidity, cognitive flexibility, and problem representation. Does sustained collaboration with generative systems expand a creator's conceptual repertoire, and if so, through what mechanisms?

- Symbiosis relationships suggest the most profound neurocognitive implications, potentially enabling hybrid cognitive capabilities that transcend unaided human thought. These deeply integrated interactions could fundamentally alter how creativity itself is experienced and conceptualized. Though speculative, research might explore neural signatures of extended symbiotic interaction, investigating potential changes in neural plasticity and the development of specialized cognitive skills that leverage complementary human and artificial processing. Such studies could employ longitudinal designs to track cognitive adaptations as creators integrate increasingly sophisticated AI extensions into their creative practice.

6.2 Psycho-social dimensions

Beyond cognitive effects, the Extended framework allows for systematic investigation of psychological and social dimensions of human-AI creative collaboration across multiple levels.

- At the *individual level*, critical questions emerge around creative identity, agency perception, and relationship to creative outputs. Support configurations may strengthen creative confidence while preserving authorial identity, but how does this experience differ from traditional tool use? Synergy relationships introduce more complex dynamics—do creators experience tensions regarding creative ownership, or might they develop more fluid conceptions of creative identity that accommodate collaborative emergence? Symbiosis configurations potentially transform fundamental aspects of creative self-concept as boundaries between human intention and machine contribution dissolve. Longitudinal studies tracking creators' evolving relationships with AI systems could illuminate these developmental trajectories.
- At *interpersonal and societal levels*, the framework enables exploration of how different human-AI configurations reshape creative communities and cultural values. How do Extended Creativity e systems influence collaboration patterns between human creators? Do they democratize access to creative expression or potentially homogenize creative outputs? When AI assistance becomes widespread, how might public conceptions of creative value evolve? Will certain extension types be privileged over others in professional contexts, and what implications might this have for educational approaches to developing creative capabilities?

These questions extend beyond individual psychology to encompass broader cultural and ethical dimensions of AI-supported creativity. Research approaches combining phenomenological investigation of lived experience with broader cultural analysis will be essential to understanding the full implications of these emerging creative relationships. Furthermore, at the social-systemic level, network analysis could provides powerful tools for modeling interaction structures in distributed creative systems involving multiple human and artificial agents. These approaches can identify emergent patterns of coordination, information flow, and influence distribution that characterize different Extended Creativity e configurations. Visualization techniques can render these complex dynamics accessible for analysis and interpretation, building on methodologies previously applied to human collaborative networks.

# 7 Implications for design

We suggest that this framework may provide a useful basis for reflecting on how different forms of human–AI creative interaction could inform the design of Extended Creativity systems. Unlike traditional approaches focused on optimizing discrete interactions, this relational perspective highlights the evolving quality of human–AI engagement—how it develops over time, shapes user agency, and supports or constrains creative growth. Below, we outline key design implications for each relational mode.

- Support systems: designing for Support relationships requires clarity, control, and consistency. These systems function best when they clearly delineate between human intent and AI execution, helping users streamline technical tasks while maintaining full conceptual authorship. Future developments should aim at domain-specific integration, enhanced recognition of personal creative styles, and more intuitive control mechanisms. The goal is to reduce cognitive load without diluting the creator's sense of ownership or expressive identity.
- Synergy systems: synergistic relationships call for interactive environments that enable mutual influence between human and AI agents. Key design principles include conversational adaptability, memory across sessions, and multimodal interfaces that support cross-domain creativity. Importantly, these systems must balance responsiveness with transparency, offering not only suggestions but also understandable rationales that foster trust and preserve human agency in the co-creative process.
- *Symbiosis systems*: Designing for Symbiosis involves radical integration of human cognition and machine generation. These systems require real-time sensing (e.g., neural input), predictive modeling, and adaptive feedback loops that respond fluidly to user intentions. Critical design challenges include ensuring cognitive transparency, protecting neural data privacy, and allowing for dynamic reconfiguration of agency distribution. In this context, design becomes a question of shaping hybrid cognitive ecologies rather than discrete tools.
- Ethical design considerations: Rather than treating ethics as an external constraint, the framework positions it as a central axis of responsible creative system design. Ethical reflection must be embedded in design practice. This includes developing transparent attribution mechanisms, maintaining human control across all stages of the creative process, accommodating neurocognitive diversity, and ensuring that AI decision-making remains intelligible.

Drawing on these considerations, it is also possible to introduce specific design-oriented metrics for evaluating the effectiveness of Extended Creativity beyond simple productivity or output quality measures:

- Creative enhancement: How does the system influence the range, originality, and quality of creative outputs compared to unassisted human creation? Does it expand creative possibilities or merely accelerate existing approaches? This assessment should consider not just the quality of the final product but also the diversity of options explored and the novelty of solutions reached.
- Epistemic surprise: We propose that the creative agency of an Extended Creativity System can be operationalized (and eventually measured) through the epistemic surprise it elicits in a human collaborator. Drawing on the Free Energy Principle (Friston, 2010), we could define creative agency as the capacity of a system to induce model updates that reduce the observer's prediction error in novel or unexpected ways. In this view, creativity emerges not solely from generativity, but from the system's capacity to act as a surprising-yet-coherent contributor to a shared meaning space.
- Experiential quality: What is the subjective experience of creating with the system? Does it foster engagement, flow, and satisfaction, or does it create friction, frustration, or alienation?
- Relationship evolution: How does the human-AI creative relationship evolve over time? Does the system adapt appropriately to changing user needs and growing expertise? The most effective systems will grow with their users, adapting as their creative capabilities and aspirations evolve.
- Social and cultural impacts: Beyond individual metrics, we must consider how these systems affect broader creative ecosystems: Do they democratize creative expression or reinforce existing hierarchies? Do they increase cultural diversity or promote homogenization? Clearly, these questions require system-level evaluations that go beyond individual interactions.

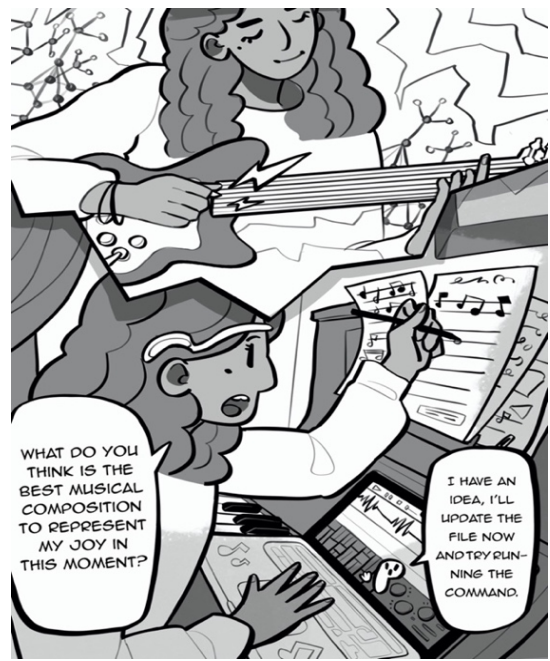

**Figure 4.** Examples of AI-human relational modes across Extended Creativity domains.

# 8  Conclusions

Examining the evolving relationship between AI and creativity reveals a key insight: designing AI systems based solely on traditional models of human creativity risks constraining their potential to explore new forms of innovation and expression—forms that may not align with human cognitive norms or aesthetic expectations. This recognition calls for a reconceptualization of creative agency, one that embraces the possibility of human–AI collaborations that *extend*, rather than replicate, the boundaries of creative practice.

Crucially, the creative value of an Extended Creativity system does not scale linearly with the level of machine autonomy or agency. Even in the Support configuration—where AI operates with limited adaptivity and causal influence—the delegation of routine, technical, or cognitively demanding tasks can amplify human creative focus, enabling more expansive ideation and deeper aesthetic exploration.

The Extended Creativity framework aims to move beyond simplistic narratives that cast AI as either a neutral tool or an existential threat to human creativity. Instead, it offers a conceptual model for understanding relational modes of creative co-agency—a shift in focus from individual capabilities to the quality and structure of human–AI interactions. The taxonomy proposed here is not a fixed classification, but a "living conceptual map", intended to evolve alongside both technological progress and experimental creative practice.

As AI systems become more sophisticated, new forms of creative collaboration will inevitably emerge—some of which may revise, or challenge key assumptions embedded in this framework. In parallel, as artists, designers, and researchers engage with these systems in novel ways, they will surface unanticipated interaction patterns that provoke theoretical refinement and practical innovation.

In the end, the most urgent question may not be whether AI-supported creativity is "real" creativity, but rather: how do these evolving human–AI relationships transform our understanding of what it means to create—and, perhaps more deeply, of what it means to relate—in a world increasingly co-shaped by natural and artificial intelligence?